\documentclass[draftcls,onecolumn]{IEEEtran}
\input{epic.sty}
\input{eepic.sty}

\newcommand{\subpostscript}[2]{
    \setlength{\epsfxsize}{#2}
    \epsfbox{#1}
}

\input{psfig}               % for including ps file
\input{epsf}                % for including eps file
\newcommand{\bthm} {\begin{thm} }
\newcommand{\ethm} {\end{thm}}
\newcommand{\blem} {\begin{lem} }
\newcommand{\elem} {\end{lem}}
\newcommand{\bcor} {\begin{cor} }
\newcommand{\ecor} {\end{cor}}

\newcommand{\beq}  {\begin{equation}}
\newcommand{\eeq}  {\end{equation}}
\newcommand{\beqa}{\begin{eqnarray}}
\newcommand{\eeqa} {\end{eqnarray}}
\newcommand{\bdis} {\begin{displaymath}}
\newcommand{\edis} {\end{displaymath}}
\newcommand{\bitem}{\begin{itemize}}
\newcommand{\eitem}{\end{itemize}}
\newcommand{\ba}  {\begin{array}}
\newcommand{\ea}  {\end{array}}

\newcommand{\bmA} {\mbox{\boldmath $A$}}
\newcommand{\bmB} {\mbox{\boldmath $B$}}
\newcommand{\bmC} {\mbox{\boldmath $C$}}

\newcommand{\bmD} {\mbox{\boldmath $D$}}
\newcommand{\bmE} {\mbox{\boldmath $E$}}
\newcommand{\bmF} {\mbox{\boldmath $F$}}
\newcommand{\bmG} {\mbox{\boldmath $G$}}

\newcommand{\bmM} {\mbox{\boldmath $M$}}
\newcommand{\bmR} {\mbox{\boldmath $R$}}

\newcommand{\bmI} {\mbox{\boldmath $I$}}
\newcommand{\bmX} {\mbox{\boldmath $X$}}

\newcommand{\bmw} {\mbox{\boldmath $w$}}

\newcommand{\bmn} {\mbox{\boldmath $n$}}

\newcommand{\bms} {\mbox{\boldmath $s$}}

\newcommand{\bmx} {\mbox{\boldmath $x$}}
\newcommand{\bmy} {\mbox{\boldmath $y$}}

\newcommand{\bmzero} {\mbox{\boldmath $0$}}

\newtheorem{theorem}{Theorem}
\newtheorem{thm}[theorem]{Theorem}
\newtheorem{lemma}{Lemma}
\newtheorem{lem}[lemma]{Lemma}

\renewcommand{\QED}{\QEDopen}

\begin{document}

\begin{titlepage}

\title{{\Large{The Impact of Noise Correlation and Channel Phase Information on the Data-Rate of
the Single-Symbol ML Decodable Distributed STBCs}}}

\author{\vspace{1cm}Zhihang~Yi and Il-Min~Kim, {\it Senior Member, IEEE}\\
\vspace{5mm}
Department of Electrical and Computer Engineering\\
Queen's University\\
Kingston, Ontario, K7L 3N6\\
Canada\\
\vspace{5mm} Email: ilmin.kim@queensu.ca}

\maketitle

\begin{center}
Submitted to {\it IEEE Trans. Inform. Theory as a Correspondence}
\end{center}
\vspace{5mm}

\begin{abstract}
Very recently, we proposed the row-monomial distributed orthogonal
space-time block codes (DOSTBCs) and showed that the row-monomial
DOSTBCs achieved approximately twice higher bandwidth efficiency
than the repetition-based cooperative strategy \cite{zhihang}.
However, we imposed two limitations on the row-monomial DOSTBCs.
The first one was that the associated matrices of the codes must
be row-monomial. The other was the assumption that the relays did
not have any channel state information (CSI) of the channels from
the source to the relays, although this CSI could be readily
obtained at the relays without any additional pilot signals or any
feedback overhead. In this paper, we first remove the row-monomial
limitation; but keep the CSI limitation. In this case, we derive
an upper bound of the data-rate of the DOSTBC and it is larger
than that of the row-monomial DOSTBCs in \cite{zhihang}. Secondly,
we abandon the CSI limitation; but keep the row-monomial
limitation. Specifically, we propose the row-monomial DOSTBCs with
channel phase information (DOSTBCs-CPI) and derive an upper bound
of the data-rate of those codes. The row-monomial DOSTBCs-CPI have
higher data-rate than the DOSTBCs and the row-monomial DOSTBCs.
Furthermore, we find the actual row-monomial DOSTBCs-CPI which
achieve the upper bound of the data-rate.

\end{abstract}


\end{titlepage}

\section{Introduction}\label{sec:intro}

In a cooperative network, the relays cooperate to help the source
transmit the information-bearing symbols to the destination. The
relay cooperation improves the performance of the network
considerably \cite{sendonaris1}--\cite{laneman2}. The cooperative
strategy of the relays is crucial and it decides the performance
of a cooperative network. A simple cooperative strategy is the
{\it repetition-based cooperative strategy} which was proposed in
\cite{laneman1} and studied in \cite{anghel1}--\cite{chen2}. This
cooperative strategy achieves the full diversity order in the
number $K$ of relays.\footnote{In this paper, unless otherwise
indicated, saying one code or one scheme achieves the full
diversity order means it achieves the full diversity in an
arbitrary signal constellation.} Furthermore, the maximum
likelihood (ML) decoding at the destination is single-symbol ML
decodable.\footnote{A code or a scheme is single-symbol ML
decodable, if its ML decoding metric can be written as a sum of
multiple terms, each of which depends on at most one transmitted
information-bearing symbol \cite{khan}.} However, the
repetition-based cooperative strategy has poor bandwidth
efficiency, since its data-rate\footnote{In this paper, the
data-rate of a cooperative strategy or a distributed space-time
code is equal to the ratio of the number of transmitted
information-bearing symbols to the number of time slots used by
the relays to transmit all these symbols.} is just $1/K$. Many
works have been devoted to improve the bandwidth efficiency of the
cooperative networks, such as the cooperative beamforming
\cite{hammerstrom}, \cite{zhihang1}, and the relay selection
\cite{zhao1}--\cite{zhihang2}. More attentions have been given to
the {\it distributed space-time codes} (DSTCs)
\cite{laneman3}--\cite{yang}. Furthermore, many practical DSTCs
have been proposed in \cite{gamal}--\cite{kiran}. Although all
those codes could improve the bandwidth efficiency, they were not
single-symbol ML decodable in general, and hence, they had much
higher decoding complexities than the repetition-based cooperative
strategy.%\footnote{The schemes proposed in \cite{yiu} would be
%single-symbol ML decodable only if the space-time code it used was
%single-symbol ML decodable. The fundamental difference between
%\cite{yiu} and this paper is that the decode-and-forward protocol
%is considered in \cite{yiu}; while we consider the
%amplify-and-forward protocol in this paper.}

Very few works have tried to propose the DSTCs achieving the
single-symbol ML decodability and the full diversity order. In
\cite{hua}, Hua {\it et al.} used the existing orthogonal designs
in cooperative networks; but they found that most codes were not
single-symbol ML decodable any more. In \cite{rajan}, Rajan {\it
et al.} used the clifford unitary weight single-symbol decodable
codes in cooperative networks. The codes were single-symbol ML
decodable only when there were four relays. Moreover, the codes
could not achieve the full diversity order in an arbitrary signal
constellation. In \cite{jing3}, Jing {\it et al.} applied the
orthogonal and quasi-orthogonal designs in cooperative networks
and they analyzed the diversity order of the codes. The authors of
\cite{jing3} claimed that the codes achieved the single-symbol ML
decodability as long as the noises at the destination were
uncorrelated. However, we noticed that the rate-$3/4$ code given
in \cite{jing3} was actually not single-symbol ML decodable,
although it generated uncorrelated noises at the destination.
Actually in this paper, we will show that, when the noises are
uncorrelated, the codes have to satisfy another constraint in
order to be single-symbol ML decodable.

Only very recently, the DSTCs achieving the single-symbol ML
decodability have been studied. In \cite{zhihang}, we proposed the
distributed orthogonal space-time block codes (DOSTBCs), and we
showed that the DOSTBCs achieved the single-symbol ML decodability
and the full diversity order. Moreover, we systematically studied
some special DOSTBCs, namely the row-monomial DOSTBCs, which
generated uncorrelated noises at the destination. Specifically, an
upper bound of the data-rate of the row-monomial DOSTBC was
derived. This upper bound suggested that the row-monomial DOSTBCs
had approximately twice higher bandwidth efficiency than the
repetition-based cooperative strategy. In \cite{zhihang}, however,
we imposed two limitations on the row-monomial DOSTBCs, in order
to simplify the analysis. The first one was that the associated
matrices of the codes must be row-monomial\footnote{A matrix is
said to be row-monomial (column-monomial) if there is at most one
non-zero entry on every row (column) of it \cite{su3}.}, which
ensured uncorrelated noises at the destination. The other was the
assumption that the relays did not have any channel state
information (CSI) of the channels from the source to the relays,
i.e. the channels of the first hop. Actually, since we assumed the
destination had the CSI of the channels from the source to the
relays and the channels from the relays to the destination in
\cite{zhihang}, the CSI of the first hop could be easily obtained
at the relays without requiring additional pilot signals or any
feedback overhead. But, it is still unknown what impact those two
limitations have on the data-rate of the codes. This has motivated
our work.

In this paper, we first abandon the row-monomial limitation; but
keep the CSI limitation. That is, we consider the DOSTBCs where
the noises at the destination are possibly correlated and the
relays do not have any CSI of the first hop. We derive an upper
bound of the data-rate of the DOSTBC and it is larger than that of
the row-monomial DOSTBC in \cite{zhihang}. This implies that the
DOSTBCs can potentially improve the bandwidth efficiency of the
cooperative network. But, like the row-monomial DOSTBCs, the
DOSTBCs may not have good bandwidth efficiency in a cooperative
network with many relays, because the upper bound of the data-rate
of the DOSTBC decreases with the number $K$ of relays. Secondly,
we remove the CSI limitation; but keep the row-monomial
limitation. Specifically, the relays know the channel phase
information (CPI) of the first hop and use this information in the
code construction. Those codes are referred to as the {\it
row-monomial DOSTBCs with CPI} (DOSTBCs-CPI). We derive an upper
bound of the data-rate of the row-monomial DOSTBC-CPI and also
find the actual codes achieving this upper bound. The upper bound
of the data-rate of the row-monomial DOSTBC-CPI is higher than
those of the DOSTBCs and the row-monomial DOSTBCs. Thus, the
row-monomial DOSTBCs-CPI have better bandwidth efficiency than the
DOSTBCs and the row-monomial DOSTBCs. Furthermore, the upper bound
of the data-rate of the row-monomial DOSTBC-CPI is independent of
the number $K$ of relays, which ensures the codes have good
bandwidth efficiency even when there are many relays.

The rest of this paper is organized as follows. Section
\ref{sec:sys} describes the cooperative network considered in this
paper. In Section \ref{sec:DOSTBC}, we remove the row-monomial
limitation; but the relays do have any CSI. Specifically, we study
the DOSTBCs and derive an upper bound of the data-rate of the
DOSTBC. In Section \ref{sec:DOSTBCcsiw}, the relays exploit the
CPI to construct the codes; but the row-monomial limitation is
maintained. Specifically, we first define the row-monomial
DOSTBCs-CPI and then derive an upper bound of the data-rate of the
row-monomial DOSTBC-CPI. We present some numerical results in
Section \ref{sec:num} and conclude this paper in Section
\ref{sec:conl}.

\emph{Notations:} Bold upper and lower letters denote matrices and
row vectors, respectively. Also, $\textrm{diag}[x_1,\cdots,x_K]$
denotes the $K\times K$ diagonal matrix with $x_1,\cdots,x_K$ on
its main diagonal; $\bmzero$ the all-zero matrix; $\bmI$ the
identity matrix; $\mathrm{det}(\cdot)$ the determinant of a
matrix; $[\cdot]_k$ the $k$-th entry of a vector;
$[\cdot]_{k_1,k_2}$ the $(k_1,k_2)$-th entry of a matrix;
$(\cdot)^{*}$ the complex conjugate; $(\cdot)^H$ the Hermitian;
$(\cdot)^T$ the transpose. Let $\bmX = [\bmx_1;\cdots;\bmx_K]$
denote the matrix with $\bmx_k$ as its $k$-th row, $1\leq k\leq
K$. For a real number $a$, $\lceil a \rceil$ denotes the ceiling
function of $a$.

%%%%%%%%%%%%%%%%%%%%%%%%%%%%%%%%%%%%%%%%%%%%%%%%%%%%%%%%%%%%%%%%%%%%

\section{System Model}\label{sec:sys}

Consider a cooperative network with one source, $K$ relays, and
one destination. Every terminal has only one antenna and is
half-duplex. Denote the channel from the source to the $k$-th
relay by $h_k$ and the channel from the $k$-th relay to the
destination by $f_k$, where $h_k$ and $f_k$ are spatially
uncorrelated complex Gaussian random variables with zero mean and
unit variance. We assume that the destination has full CSI, i.e.
it knows the instantaneous values of $h_k$ and $f_k$ by using
pilot signals; while the source has no CSI. The relays may have
partial CSI and this will be discussed in detail later.

At the beginning, the source transmits $N$ complex-valued
information-bearing symbols over $N$ consecutive time
slots.\footnote{If the information-bearing symbols are
real-valued, one can use the rate-one generalized real orthogonal
design proposed by \cite{tarokh} in the cooperative networks
without any changes. The codes achieve the single-symbol ML
decodability and the full diversity order \cite{hua}. Therefore,
we focus on the complex-valued symbols in this paper.} Let
$\bms=[s_1,\cdots,s_N]$ denote the information-bearing symbol
vector transmitted from the source, where the power of $s_n$ is
$E_s$. Assume the coherence time of $h_k$ is larger than $N$; then
the received signal vector $\bmy_k$ at the $k$-th relay is $
\bmy_k = h_k\bms+\bmn_k$, where $\bmn_k =
[n_{k,1},\cdots,n_{k,N}]$ is the additive noise at the $k$-th
relay and it is uncorrelated complex Gaussian with zero mean and
identity covariance matrix. All the relays are working in the
amplify-and-forward mode and the amplifying coefficient $\rho$ is
$\sqrt{E_r/(1+E_s)}$ for every relay, where $E_r$ is the
transmission power at every relay.\footnote{We set
$\rho=\sqrt{E_r/(1+E_s)}$ as in many previous publications
including \cite{nabar,jing1,rajan,jing3}. This ensures the average
transmission power of every relay is $E_r$ in a long term.} Based
on the received signal vector $\bmy_k$, the $k$-th relay produces
a transmitted signal vector and forwards it to the destination.

Firstly, we present the system model of the DOSTBCs, which will be
studied in Section \ref{sec:DOSTBC}. We assume that the $k$-th
relay has no CSI of the first hop. This can be true when the
relays do not have any channel estimation devices due to strict
power and/or size constraints.\footnote{Even when the relays can
not estimate the channels, the destination is still able to obtain
the full CSI. This is because the destination usually dose not
have any power or size limitation, and hence, it can be equipped
with sophisticated channel estimation devices. Furthermore,
although the relays can not estimate the channels, they can
forward the pilot signals from the source to the destination, and
they can transmit their own pilot signals to the destination.
Based on those pilot signals, the destination is able to obtain
the full CSI, which has been discussed in \cite{zhihang2} and
\cite{hua}.} Then the $k$-th relay produces the transmitted signal
vector $\bmx^D_{k}$ as follows:
\begin{eqnarray}\label{eqn:bmxk}
\bmx^D_{k} &=& \rho(\bmy_k\bmA_k+\bmy_k^{*}\bmB_k)\nonumber\\
&=&\rho h_k \bms \bmA_k +\rho h_k^{*}
\bms^{*}\bmB_k+\rho\bmn_k\bmA_k+\rho\bmn_k^{*}\bmB_k.
\end{eqnarray}
The matrices $\bmA_k$ and $\bmB_k$ are called the associated
matrices. They have the dimension of $N\times T$ and their
properties will be discussed in detail later. Assume the coherence
time of $f_k$ is larger than $T$. The received signal vector at
the destination is given by
\begin{eqnarray}
\bmy_D &=& \sum_{k=1}^K f_k\bmx^D_k + \bmn_d\nonumber\\
&=&\sum_{k=1}^K(\rho f_k h_k \bms \bmA_k +\rho f_k h_k^{*}
\bms^{*}\bmB_k)+\sum_{k=1}^K(\rho f_k\bmn_k\bmA_k+\rho
f_k\bmn_k^{*}\bmB_k)+\bmn_d, \label{eqn:ydfull2}
\end{eqnarray}
where $\bmn_d = [n_{d,1},\cdots,n_{d,T}]$ is the additive noise at
the destination and it is uncorrelated complex Gaussian with zero
mean and identity covariance matrix.\footnote{We assume that there
is no direct link between the source and destination. The same
assumption has been made in many previous publications
\cite{yiu,jing1,li,jing3}. Furthermore, perfect synchronization
among the relays is assumed as in \cite{gamal,yiu,jing1}, and
\cite{hua}--\cite{jing3}. Although synchronization is a critical
issue for the practical implementation of cooperative networks, it
is beyond the scope of this paper.} Define $\bmw_D$, $\bmX_D$, and
$\bmn_D$ as follows:
\begin{eqnarray}
\bmw_D&=&[\rho f_1,\cdots,\rho f_K]\label{eqn:w}\\
\bmX_D&=&[h_1\bms\bmA_1+h_1^{*}\bms^{*}\bmB_1;\cdots;h_K\bms\bmA_K+h_K^{*}\bms^{*}\bmB_K]\label{eqn:X}\\
\bmn_D&=&\sum_{k=1}^K(\rho f_k\bmn_k\bmA_k+\rho
f_k\bmn_k^{*}\bmB_k)+\bmn_d;\label{eqn:n}
\end{eqnarray}
then we can rewrite (\ref{eqn:ydfull2}) in the following way
\begin{equation}
\bmy_D = \bmw_D \bmX_D +\bmn_D.\label{eqn:yd}
\end{equation}
%where $\bmw$ is the equivalent channel, $\bmX$ is the code matrix,
%and $\bmn$ is the equivalent noise.
Furthermore, from (\ref{eqn:n}), it is easy to see that the mean
of $\bmn_D$ is zero and the covariance matrix $\bmR$ of $\bmn_D$
is given by
\begin{eqnarray}
\bmR &=& \sum_{k=1}^K \left(|\rho f_k|^2\left(\bmA_k^H\bmA_k
+\bmB_k^H\bmB_k\right)\right)+\bmI.\label{eqn:R}
\end{eqnarray}

Secondly, we present another system model, which is for the
row-monomial DOSTBCs-CPI studied in Section \ref{sec:DOSTBCcsiw}.
We assume that there is no strict power or size constraint on the
relays and the relays can obtain partial CSI of the first hop by
the equipped channel estimation devices. Specifically, we assume
the $k$-th relay has the CPI of the first hop, i.e. it knows the
phase $\theta_k$ of the channel coefficient $h_k$.\footnote{In
this paper, we assume that the relays can estimate $\theta_k$
without any errors as in \cite{hua}--\cite{jing3}. It will be
interesting to study the scenario when the relays do not have
perfect estimations of $\theta_k$; but it is beyond the scope of
this paper.} Note that this assumption does not imply more pilot
signals compared to the assumption that relays have no CSI of the
first hop. Actually, in order to make the destination have full
CSI, the relays always need to forward the pilot signals from the
source to the relays. Furthermore, the relays always need to
transmit their own pilot signals to the destination. Therefore,
the same amount of pilot signals is needed in all circumstances.
Furthermore, the assumption that the relays have the CPI of the
first hop does not imply any feedback overhead, because the relays
do not need to have any CSI of the channels from themselves to the
destination.

Based on the CPI, the $k$-th relay first obtains $\bmy^C_k$ by
$\bmy^C_k = e^{-j\theta_k}\bmy_k$ and then builds the transmitted
signal vector $\bmx^C_{k}$ as
\begin{eqnarray}
\bmx_{k}^C &=& \rho(\bmy^C_k\bmA_k+\bmy_k^{C*}\bmB_k)\nonumber\\
&=&\rho |h_k| \bms \bmA_k +\rho |h_k| \bms^{*}\bmB_k+\rho
e^{-j\theta_k}\bmn_k\bmA_k+\rho e^{j\theta_k}\bmn_k^{*}\bmB_k.
\end{eqnarray}
Consequently, the received signal vector at the destination is
given by
\begin{equation}
\bmy_C = \bmw_C \bmX_C +\bmn_C\label{eqn:yc},
\end{equation}
where
\begin{eqnarray}
\bmw_C&=&[\rho f_1|h_1|,\cdots,\rho f_K|h_K|]\\
\bmX_C&=&[\bms\bmA_1+\bms^{*}\bmB_1;\cdots;\bms\bmA_K+\bms^{*}\bmB_K]\label{eqn:Xcsi}\\
\bmn_C&=&\sum_{k=1}^K(\rho f_ke^{-j\theta_k}\bmn_k\bmA_k+\rho
f_ke^{j\theta_k}\bmn_k^{*}\bmB_k)+\bmn_d.\label{eqn:ncsi}
\end{eqnarray}
From (\ref{eqn:ncsi}), it is easy to see that the mean of $\bmn_C$
is zero and the covariance matrix $\bmR$ of $\bmn_C$ is still
given by (\ref{eqn:R}).

%%%%%%%%%%%%%%%%%%%%%%%%%%%%%%%%%%%%%%%%%%%%%%%%%%%%%%%%%%%%%%%%%%%%%

\section{Distributed Orthogonal Space-Time Block
Codes}\label{sec:DOSTBC}

In this section, we abandon the row-monomial limitation, which was
adopted in the construction of the row-monomial DOSTBCs in
\cite{zhihang}. Thus, the codes possibly generate correlated
noises at the destination. However, we still keep the CSI
limitation, i.e. the relays do not have any CSI. It is easy to see
that such codes are just the DOSTBCs proposed in \cite{zhihang},
whose definition is as follows.

\vspace{3mm} {\it Definition 1:} A $K\times T$ code matrix
$\bmX_D$ is called a DOSTBC in variables $s_1,\cdots,s_N$ if the
following two conditions are satisfied:
\begin{description}
    \item[D1.1)] The entries of $\bmX_D$ are 0, $\pm h_k s_n$,
    $\pm h_k^{*}s_n^{*}$, or multiples of these indeterminates by $\textbf{j}$, where $\textbf{j} =
    \sqrt{-1}$.
    \item[D1.2)] The matrix $\bmX_D$ satisfies the following equality
    \begin{eqnarray}\label{eqn:orth}
    \bmX_D\bmR^{-1}\bmX_D^H &=& |s_1|^2 \bmD_1 +\cdots +|s_N|^2
    \bmD_N,
    \end{eqnarray}
    where $\bmD_n={\mathrm{diag}}[|h_1|^2D_{n,1},\cdots,|h_K|^2D_{n,K}]$
    and $D_{n,1},\cdots,D_{n,K}$ are non-zero.
\end{description}
\vspace{3mm}

In \cite{zhihang}, it has been shown the DOSTBCs are single-symbol
ML decodable and achieve the full diversity order $K$. However,
the bandwidth efficiency of the DOSTBCs has not been analyzed in
\cite{zhihang}. Thus, it is still unknown if removing the
row-monomial limitation can improve the bandwidth efficiency or
not. In order to answer this question, we derive an upper bound of
the data-rate of the DOSTBC in the following. To this end, one may
think of redefining $\bmw_D$ and $\bmX_D$ as follows
\begin{eqnarray}
\bmw_D&=&[\rho f_1|h_1|,\cdots,\rho f_K|h_K|]\label{eqn:neww}\\
\bmX_D&=&[\bms \tilde{\bmA}_1+\bms^{*}\tilde{\bmB}_1 ;\cdots;\bms
\tilde{\bmA}_K+\bms^{*}\tilde{\bmB}_K],\label{eqn:newx}
\end{eqnarray}
where $\tilde{\bmA}_k = e^{j\theta_k}\bmA_k$ and $\tilde{\bmB}_k =
e^{-j\theta_k}\bmB_k$. Then $\bmX_D$ can be seen as the
generalized orthogonal designs, and hence, the results in
\cite{wang} may be directly used. Actually, this method will make
the analysis more complicated. Note that the new associated
matrices $\tilde{\bmA}_k$ and $\tilde{\bmB}_k$ have a fundamental
difference with the associated matrices of the generalized
orthogonal design in \cite{wang}. That is, $\tilde{\bmA}_k$ and
$\tilde{\bmB}_k$ contain $\theta_k$, which is a random variable.
Due to this reason, it is very hard to find the properties of
$\tilde{\bmA}_k$ and $\tilde{\bmB}_k$ by using the results in
\cite{wang}, and hence, it is very complicated to derive an upper
bound by using (\ref{eqn:neww}) and (\ref{eqn:newx}). Instead of
this approach, in this paper, we define $\bmw_D$ and $\bmX_D$ as
in (\ref{eqn:w}) and (\ref{eqn:X}), respectively, and derive an
upper bound of the data-rate by analyzing the properties of
$\bmA_k$ and $\bmB_k$. Some fundamental properties of $\bmA_k$ and
$\bmB_k$ are given in the following lemma at first.

\begin{lem}\label{lem:propertyAB}
If a DOSTBC $\bmX_D$ in variables $s_1,\cdots,s_N$ exists, its
associated matrices $\bmA_k$ and $\bmB_k$ are column-monomial.
Furthermore, the orthogonal condition (\ref{eqn:orth}) on $\bmX_D$
holds if and only if
\begin{eqnarray}
\bmA_{k_1}\bmR^{-1}\bmA_{k_2}^H &=& \bmzero, \hspace{2cm} k_1 \neq
k_2 \label{eqn:cond1}
\\
\bmB_{k_1}\bmR^{-1}\bmB_{k_2}^H &=& \bmzero, \hspace{2cm} k_1
\neq k_2 \label{eqn:cond2}\\
\bmA_{k_1} \bmR^{-1} \bmB_{k_2}^H + \bmB_{k_2}^{*}\bmR^{-1}
\bmA_{k_1}^T &=& \bmzero, \hspace{2cm} \label{eqn:cond3}\\
\bmB_{k_1}\bmR^{-1}\bmA_{k_2}^H +
\bmA_{k_2}^*\bmR^{-1}\bmB_{k_1}^T
&=& \bmzero, \hspace{2cm} \label{eqn:cond4}\\
\bmA_k \bmR^{-1} \bmA_k^H + \bmB^{*}_k \bmR^{-1} \bmB^T_ k&=&
{\mathrm{diag}}[D_{1,k},\cdots,D_{N,k}].\label{eqn:cond5}
\end{eqnarray}
\end{lem}
\begin{proof}
By following the proof of Property 3.2 in \cite{su3}, it is very
easy to show that $\bmA_k$ and $\bmB_k$ are column-monomial.
Furthermore, by following the proof of Lemma 1 in \cite{zhihang}
and the proof of Proposition 1 in \cite{wang}, it is not hard to
show (\ref{eqn:cond1})--(\ref{eqn:cond5}).
\end{proof}

Lemma \ref{lem:propertyAB} gives us some fundamental properties of
$\bmA_k$ and $\bmB_k$. But, due to the existence of $\bmR^{-1}$,
we can not obtain an upper bound by using the conditions
(\ref{eqn:cond1})--(\ref{eqn:cond5}) directly. Therefore, we
simplify those conditions in the following theorem by eliminating
$\bmR^{-1}$.

%Furthermore, similar to the definition of generalized complex
%orthogonal design, we constrain that $\bmX$ does not contain the
%linear combination of $\pm h_k s_n$ and $\pm h_k^{*} s_n^{*}$.

\begin{thm}\label{thm:necessary}
If a DOSTBC $\bmX_D$ in variables $s_1,\cdots,s_N$ exists, we have
\begin{eqnarray}\label{eqn:orthE}
\bmX_D\bmX_D^H&=&|s_1|^2\bmE_1 + \cdots+ |s_N|^2\bmE_N,
\end{eqnarray}
where $\bmE_n = {\mathrm{diag}}[|h_1|^2 E_{n,1},\cdots,|h_K|^2
E_{n,K}] $ and $E_{n,1},\cdots,E_{n,K}$ are strictly positive.
Equivalently, the associated matrices $\bmA_k$ and $\bmB_k$
satisfy the following conditions
\begin{eqnarray}
\bmA_{k_1}\bmA_{k_2}^H &=& \bmzero, \hspace{2cm}  k_1 \neq k_2
\label{eqn:Econd1}
\\
\bmB_{k_1}\bmB_{k_2}^H &=& \bmzero, \hspace{2cm}  k_1
\neq k_2 \label{eqn:Econd2}\\
\bmA_{k_1}  \bmB_{k_2}^H + \bmB_{k_2}^{*}
\bmA_{k_1}^T &=& \bmzero  \label{eqn:Econd3}\\
\bmB_{k_1}\bmA_{k_2}^H + \bmA_{k_2}^*\bmB_{k_1}^T
&=& \bmzero\label{eqn:Econd4}\\
\bmA_k \bmA_k^H + \bmB^{*}_k \bmB^T_ k&=&
{\mathrm{diag}}[E_{1,k},\cdots,E_{N,k}].\label{eqn:Econd5}
\end{eqnarray}
\end{thm}
\begin{proof}
See Appendix A.
\end{proof}

After comparing Theorem \ref{thm:necessary} and the definition of
the generalized orthogonal design in \cite{wang}, it seems that
the DOSTBCs are in a subset of the generalized orthogonal design.
However, note that there is a fundamental difference between the
DOSTBCs and the generalized orthogonal design. That is, the code
matrix $\bmX_D$ of a DOSTBC contains the channel coefficients
$h_k$. Actually, this fundamental difference explains why
$\tilde{\bmA}_k$ and $\tilde{\bmB}_k$ in (\ref{eqn:newx}) contain
$\theta_k$. Furthermore, this fundamental difference induces the
conditions (\ref{eqn:Econd1}) and (\ref{eqn:Econd2}). Those two
conditions help derive an upper bound of the data-rate of the
DOSTBC in the following theorem.

\begin{thm}\label{thm:upperDOSTBC}
If a DOSTBC $\bmX_D$ in variables $s_1,\cdots,s_N$ exists, its
data-rate ${\cal{R}}_D$ satisfies the following inequality:
\begin{equation}\label{eqn:rateDOSTBC}
{\cal{R}}_D=\frac{N}{T} \leq \frac{N}{\lceil \frac{NK}{2} \rceil}.
\end{equation}
%where $\lceil (NK)/2 \rceil$ denotes the ceiling function of
%$(NK)/2 $.
\end{thm}
\begin{proof}
See Appendix B.
\end{proof}

Theorem \ref{thm:upperDOSTBC} suggests that the DOSTBCs have
approximately twice higher bandwidth efficiency than the
repetition-based cooperative strategy. Furthermore, it is worthy
of addressing that the DOSTBCs have the same decoding complexity
and diversity order as the repetition-based cooperative strategy.
On the other hand, when $N$ and $K$ are both even, the upper bound
of the data-rate of the DOSTBC is exactly the same as that of the
row-monomial DOSTBC proposed in \cite{zhihang}. Therefore, one can
use the systematic construction method developed in Section V-A of
\cite{zhihang} to build the DOSTBCs achieving the upper bound of
(\ref{eqn:rateDOSTBC}) for this case. For the other cases when $N$
and/or $K$ are odd, the upper bound of the data-rate of the DOSTBC
is larger than that of the row-monomial DOSTBC. Unfortunately, we
have not found any DOSTBCs achieving the upper bound of
(\ref{eqn:rateDOSTBC}) for those cases.

On the other hand, like the row-monomial DOSTBCs, the DOSTBCs may
have poor bandwidth efficiency, when there are many relays. This
is because the upper bound (\ref{eqn:rateDOSTBC}) decreases with
the number $K$ of relays. This problem can be solved very well
when the relays can exploit the CPI to construct the codes, which
will be shown in the next section.

%%%%%%%%%%%%%%%%%%%%%%%%%%%%%%%%%%%%%%%%%%%%%%%%%%%%%%%%%%%%%%%%%%%%%%%%%%

\section{Row-Monomial Distributed Orthogonal Space-Time Block
Codes with Channel Phase Information}\label{sec:DOSTBCcsiw}

In this section,  we remove the CSI limitation and assume the
relays use the CPI to construct the codes. But, we still keep the
row-monomial limitation, in order to facilitate the analysis.
Therefore, we define the row-monomial DOSTBCs-CPI in the following
way.

\vspace{3mm}
 {\it Definition 2:} A $K\times T$ code matrix $\bmX_C$ is
called a row-monomial DOSTBC-CPI in variables $s_1,\cdots,s_N$ if
it satisfies D1.1 in Definition $1$ and the following equality
 \begin{eqnarray}\label{eqn:orthcsi}
    \bmX_C\bmR^{-1}\bmX_C^H &=& |s_1|^2 \bmF_1 +\cdots +|s_N|^2
    \bmF_N,
    \end{eqnarray}
    where $\bmF_n={\mathrm{diag}}[F_{n,1},\cdots,F_{n,K}]$
    and $F_{n,1},\cdots,F_{n,K}$ are non-zero. Furthermore, it associated matrices $\bmA_k$ and $\bmB_k$, $1\leq
k \leq K$, are all row-monomial. \vspace{3mm}

It is easy to check that the row-monomial DOSTBCs-CPI are
single-symbol ML decodable. By using the technique in
\cite{zhihang} and \cite{elia}, it can be shown that the
row-monomial DOSTBCs-CPI also achieve the full diversity order.
Furthermore, by following the proof of Lemma \ref{lem:propertyAB}
and Theorem \ref{thm:necessary}, it is not hard to show that a
row-monomial DOSTBC-CPI $\bmX_C$ satisfies the follow equality
\begin{eqnarray}\label{eqn:orthEcsi}
\bmX_C\bmX_C^H&=&|s_1|^2\bmG_1 + \cdots+ |s_N|^2\bmG_N,
\end{eqnarray}
where $\bmG_n = {\mathrm{diag}}[G_{n,1},\cdots,G_{n,K}]$ and
$G_{n,1},\cdots,G_{n,K}$ are strictly positive. Note that the code
matrix $\bmX_C$ of a row-monomial DOSTBC-CPI does not contain any
channel coefficients. By comparing (\ref{eqn:orthEcsi}) and the
definition of the generalized orthogonal design, we notice that a
row-monomial DOSTBC-CPI must be a generalized orthogonal design.
%Therefore, its associated matrices $\bmA_k$ and $\bmB_k$ must
%satisfy the following conditions
%\begin{eqnarray}
%\bmA_{k_1}\bmA_{k_2}^H +
%\bmB_{k_2}^*\bmB_{k_1}^T&=&\delta_{k_1,k_2}{\mathrm{diag}}[E_{1,k_1},\cdots,E_{N,k_1}]\label{eqn:aabb}\\
%\bmA_{k_1}\bmB_{k_2}^H + \bmB_{k_2}^*\bmA_{k_1}^T &=& \bmzero \label{eqn:abba}\\
%\bmB_{k_1}\bmA_{k_2}^H + \bmA_{k_2}^*\bmB_{k_1}^T &=&
%\bmzero\label{eqn:baab}.
%\end{eqnarray}
All the analysis of the generalized orthogonal design in
\cite{wang} are valid for the row-monomial DOSTBCs-CPI. In
particular, when $K=2$, the data-rate of the row-monomial
DOSTBC-CPI can be as large as one by using the Alamouti code
proposed in \cite{alamouti}; when $K>2$, the data-rate of the
row-monomial DOSTBC-CPI is upper-bounded $4/5$, which is the upper
bound of the data-rate of the generalized orthogonal design
\cite{wang}.

Actually, the row-monomial DOSTBCs-CPI have some unique properties
which the generalized orthogonal design does not have. Those
unique properties help find a tighter upper bound of the data-rate
of the row-monomial DOSTBC-CPI. To this end, we first have the
following theorem.

%\begin{thm}\label{thm:XRX}
%If a row-monomial DOSTBC-CPI $\bmX_C$ in variables
%$s_1,\cdots,s_N$ exists, we have
%\begin{eqnarray}\label{eqn:orthF}
%\bmX_C\bmR\bmX_C^H&=&|s_1|^2\bmF_1 + \cdots+ |s_N|^2\bmF_N,
%\end{eqnarray}
%where $\bmF_n= {\mathrm{diag}}[F_{n,1},\cdots,F_{n,K}]$.

%\end{thm}
%\begin{proof}
%See Appendix C.
%\end{proof}

%Furthermore, a row-monomial DOSTBC can be partitioned into several
%sub-matrices and each of them is still a row-monomial DOSTBC.

\begin{thm}\label{thm:partition}
Assume $\bmX_C$ is a DSTC in variables $s_1,\cdots,s_N$, i.e.
every row of $\bmX_C$ contains the information-bearing symbols
$s_1,\cdots,s_N$. Moreover, assume that the noise covariance
matrix $\bmR$ of $\bmX_C$ is diagonal. After proper column
permutations, we can partition $\bmR^{-1}$ into $\bmR^{-1} =
{\mathrm{diag}}[\bmR_1,\bmR_2,\cdots,\bmR_W]$ such that the main
diagonal entries of $\bmR_w$ are all equal to $R_w$ and $R_{i}\neq
R_{j}$ for $i\neq j$. After the same column permutations, we can
partition $\bmX_C$ into $\bmX_C =[\bmX_{C1},\cdots,\bmX_{CW}]$.
Let $\tilde{\bmX}_{Cw}$ denote all the non-zero rows in
$\bmX_{Cw}$. Assume that $\tilde{\bmX}_{Cw}$ contains $N_w$
different information-bearing symbols and they are
$s^w_1,\cdots,s^w_{N_w}$.\footnote{Note that
$s^w_1,\cdots,s^w_{N_w}$ are all from the set
$\bms=[s_1,\cdots,s_N]$.} Then $\bmX_C$ is a row-monomial
DOSTBC-CPI if and only if every sub-matrix $\tilde{\bmX}_{Cw}$ is
a row-monomial DOSTBC-CPI in variables $s^w_1,\cdots,s^w_{N_w}$.
\end{thm}
\begin{proof}
See Appendix C.
\end{proof}

Theorem \ref{thm:partition} means that, when a DSTC $\bmX_C$
generates uncorrelated noises at the destination, the code is
single-symbol ML decodable as long as it can be partitioned into
several single-symbol ML decodable codes.\footnote{For the
rate-$3/4$ code in \cite{jing3}, it generates uncorrelated noises
at the destination; but the main diagonal entries of $\bmR$ are
all different. If we partition the rate-$3/4$ code by the way
presented in Theorem \ref{thm:partition}, we will see that every
sub-matrix $\tilde{\bmX}_{Cw}$ is actually a column vector with
more than one non-zero entries. Thus, $\tilde{\bmX}_{Cw}$ can not
be a row-monomial DOSTBC-CPI, and hence, it is not single-symbol
ML decodable. By Theorem \ref{thm:partition}, the rate-$3/4$ code
can not be a row-monomial DOSTBC-CPI either and it is not
single-symbol ML decodable.} Furthermore, Theorem
\ref{thm:partition} is crucial to derive an upper bound of the
data-rate of the row-monomial DOSTBC-CPI. This is because it
enables us to analyze the data-rate of every individual sub-matrix
$\tilde{\bmX}_{Cw}$ instead of $\bmX_C$ itself. When
$\tilde{\bmX}_{Cw}$ has one or two rows, it is easy to see that
its data-rate can be as large as one. When $\tilde{\bmX}_{Cw}$ has
more than two rows, the following theorem shows that the data-rate
of $\tilde{\bmX}_{Cw}$ is exactly $1/2$.

\begin{thm}\label{thm:row3}
Assume $\bmX_C$ is a row-monomial DOSTBC-CPI and its noise
covariance matrix is $\bmR$. By proper column permutations, we can
partition $\bmR^{-1}$ into $\bmR^{-1} =
{\mathrm{diag}}[\bmR_1,\bmR_2,\cdots,\bmR_W]$ such that the main
diagonal entries of $\bmR_w$ are all equal to $R_w$ and $R_{i}\neq
R_{j}$ for $i\neq j$. By the same column permutations, we can
partition $\bmX_C$ into $\bmX_C =[\bmX_{C1},\cdots,\bmX_{CW}]$.
Let $\tilde{\bmX}_{Cw}$ denote all the non-zero rows in
$\bmX_{Cw}$ and assume the dimension of $\tilde{\bmX}_{Cw}$ is
$K_w\times T_w$. Then the data-rate of $\tilde{\bmX}_{Cw}$ is
exactly $1/2$ when $K_w>2$.
\end{thm}
\begin{proof}
See Appendix D.
\end{proof}

Based on Theorems \ref{thm:partition} and \ref{thm:row3}, we
derive an upper bound of the data-rate of the row-monomial
DOSTBC-CPI in the following theorem.
\begin{thm}\label{thm:rateDOSTBCscsi}
When $K>2$, the data-rate ${\cal{R}}_C$ of the row-monomial
DOSTBC-CPI satisfies the following inequality
\begin{equation}\label{eqn:rateDOSTBCcsi}
{\cal{R}}_C=\frac{N}{T} \leq \frac{1}{2}.
\end{equation}
\end{thm}
\begin{proof}
See Appendix E.
\end{proof}

We notice that the data-rate of the row-monomial DOSTBC-CPI is
independent of the number $K$ of relays. Thus, the row-monomial
DOSTBCs-CPI have good bandwidth efficiency even in a cooperative
network with many relays. Furthermore, compared to the
row-monomial DOSTBCs and the DOSTBCs, the row-monomial DOSTBCs-CPI
improve bandwidth efficiency considerably, especially when the
cooperative network has a large number of relays. The improvement
is mainly because the relays exploit the CPI to construct the
codes. As we have seen, because the code matrix $\bmX_D$ of a
DOSTBC contains the channel coefficient $h_k$, the conditions
(\ref{eqn:Econd1}) and (\ref{eqn:Econd2}) are induced. Those two
conditions severely constrain the data-rate of the DOSTBC. On the
other hand, by exploiting the CPI, the code matrix $\bmX_C$ of a
row-monomial DOSTBC-CPI does not have any channel coefficients.
Thus, the conditions (\ref{eqn:Econd1}) and (\ref{eqn:Econd2}) are
not induced, and the data-rate is greatly improved. Furthermore,
recall that exploiting the CPI at the relays does not increase the
pilot signals or require any feedback overhead. Compared to the
DOSTBCs, the only extra cost of the row-monomial DOSTBCs-CPI is
that the relays should be equipped with some channel estimation
devices to estimate $\theta_k$.

Interestingly, the row-monomial DOSTBCs-CPI achieving the upper
bound $1/2$ are easy to construct and they are given in the
following theorem.

\begin{thm}
The rate-halving codes developed in \cite{tarokh} can be used as
the row-monomial DOSTBCs-CPI achieving the upper bound $1/2$ of
the data-rate.
\end{thm}
\begin{proof}
It is easy to check that the rate-halving codes satisfy Definition
2 and they always achieve the data-rate $1/2$.
\end{proof}

As an example, when $N=4$ and $K=4$, the row-monomial DOSTBC-CPI
achieving the upper bound $1/2$ is given as follows:
\begin{equation}
\bmX_C=\left[%
\begin{array}{cccccccc}
  s_1 & -s_2 & -s_3 & -s_4 & s_1^* & -s_2^* & -s_3^* & -s_4^* \\
  s_2 & s_1 & s_4 & -s_3 & s_2^* & s_1^* & s_4^* & -s_3^*\\
  s_3 & -s_4 & s_1 & s_2 & s_3^* & -s_4^* & s_1^* & s_2^* \\
  s_4 & s_3 & -s_2 & s_1 & s_4^* & s_3^* & -s_2^* & s_1^* \\
\end{array}%
\right].
\end{equation}

%%%%%%%%%%%%%%%%%%%%%%%%%%%%%%%%%%%%%%%%%%%%%%%%%%%%%%%%%%%%%%%%%%%%%

\section{Numerical Results}\label{sec:num}

In this section, we present some numerical results to demonstrate
the performance of the DOSTBCs and the row-monomial DOSTBCs-CPI.
In our simulation, we define the average signal to noise ratio
(SNR) per bit as the ratio of $E_r$ to the logarithm of the size
of the modulation scheme. Furthermore, we adopt the power
allocation proposed in \cite{jing1}, i.e. $E_s=KE_r$.

\begin{figure}
\begin{center}
\subpostscript{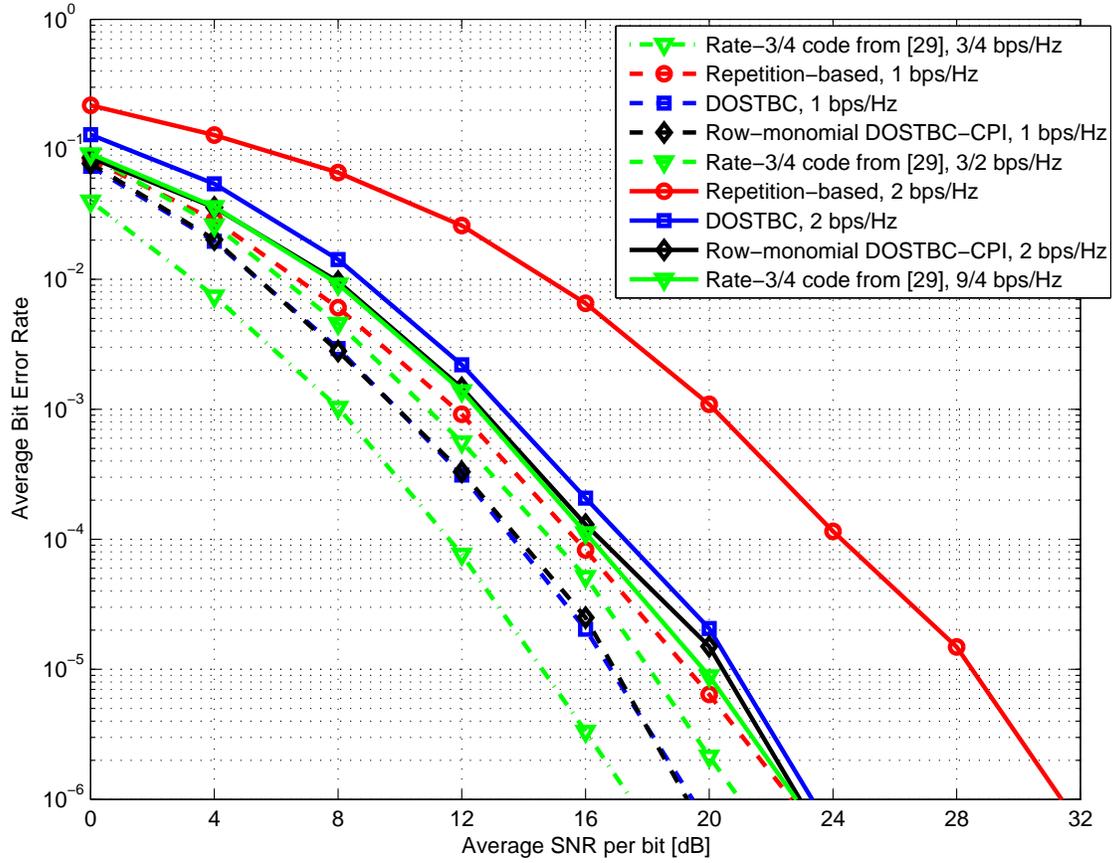}{0.9\textwidth}
\end{center}
\caption{Comparison of the rate-$3/4$ code from \cite{jing3}, the
DOSTBCs, the row-monomial DOSTBCs-CPI, and the repetition-based
cooperative strategy, $N=4$, $K=4$.} \label{fig:n4k4_halving}
\end{figure}

\begin{figure}
\begin{center}
\subpostscript{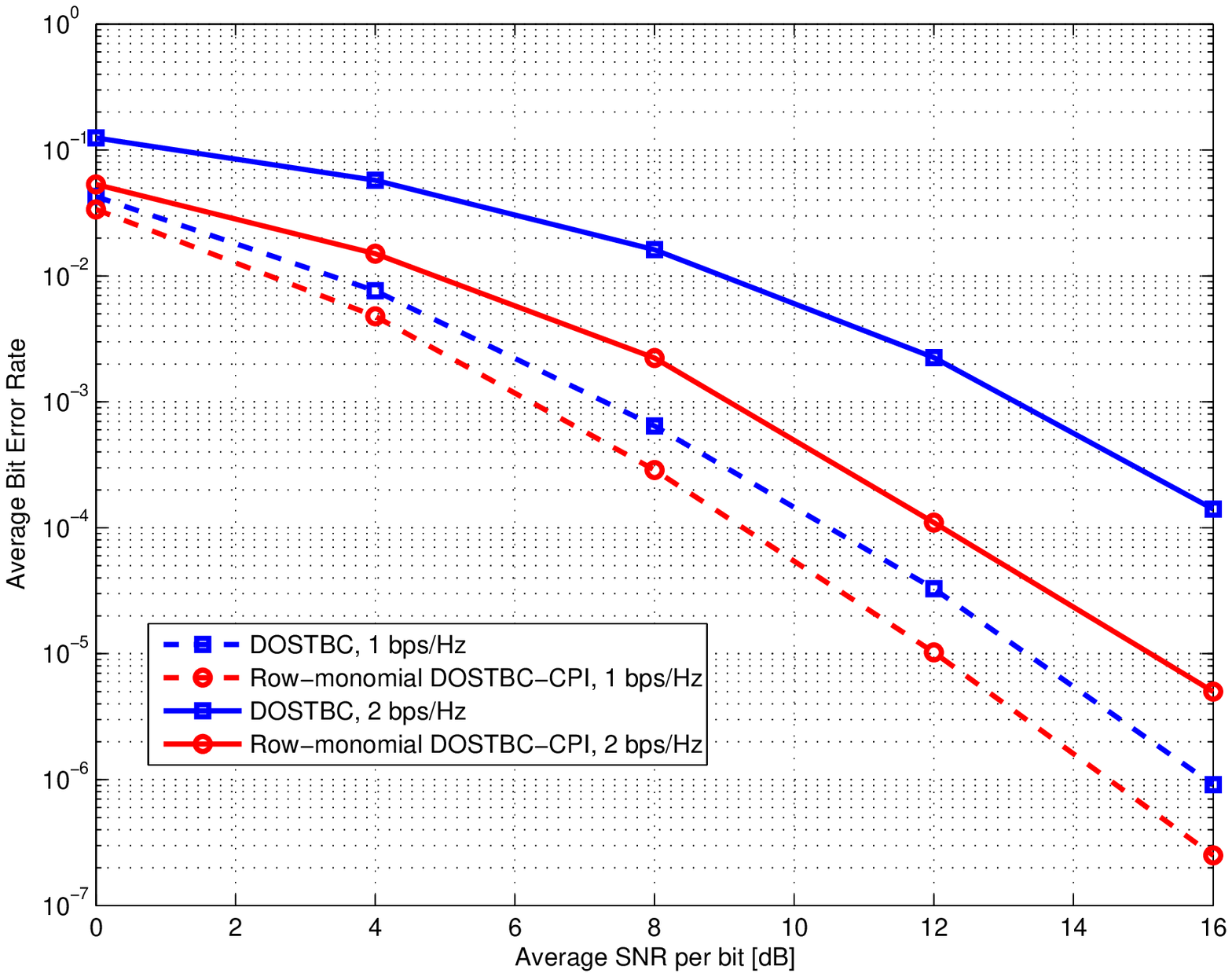}{0.9\textwidth}
\end{center}
\caption{Comparison of the DOSTBCs and the row-monomial
DOSTBCs-CPI, $N=8$, $K=6$.} \label{fig:n8k6_halving}
\end{figure}

In Fig.\ \ref{fig:n4k4_halving}, we let $N=4$ and $K=4$. For this
case, we see that the average bit error rate (BER) performance of
the DOSTBCs and the row-monomial DOSTBCs-CPI is much better than
that of the repetition-based cooperative strategy, especially when
the bandwidth efficiency is $2$ bps/Hz. The DOSTBCs and the
row-monomial DOSTBCs-CPI have almost the same performance. This is
because, when $N=4$ and $K=4$, the DOSTBCs and the row-monomial
DOSTBCs-CPI have the same data-rate $1/2$. Fig.\
\ref{fig:n4k4_halving} also demonstrates that the performance of
the DOSTBCs and the row-monomial DOSTBCs-CPI is slightly worse
than that of the rate-$3/4$ code proposed in \cite{jing3}. But,
note that the rate-$3/4$ code is not single-symbol ML decodable,
and hence, its decoding complexity is much higher than that of the
DOSTBCs and the row-monomial DOSTBCs-CPI. In Fig.
\ref{fig:n8k6_halving}, we set $N=8$ and $K=6$. For this case, the
average BER performance of the row-monomial DOSTBCs-CPI is now
much better than that of the DOSTBCs. This is because, when $N=8$
and $K=6$, the data-rate of the row-monomial DOSTBC-CPI is still
$1/2$; while the data-rate of the DOSTBC becomes $1/3$.

\section{Conclusion and Future Work}\label{sec:conl}

In the first part of this paper, we consider the DOSTBCs, where
the noises at the destination are possibly correlated and the
relays have no CSI of the first hop. An upper bound of the
data-rate of the DOSTBC is derived. When $N$ and $K$ are both
even, the upper bond of the data-rate of the DOSTBC is exactly the
same as that of the row-monomial DOSTBC in \cite{zhihang}. When
$N$ and/or $K$ are odd, the upper bound of the data-rate of the
DOSTBC is larger than that of the row-monomial DOSTBC, which means
the DOSTBCs can potentially improve the bandwidth efficiency.
However, we notice that, like the row-monomial DOSTBCs, the
DOSTBCs may not have good bandwidth efficiency in a cooperative
network with many relays, because the upper-bound of the data-rate
of the DOSTBC decreases with the number $K$ of the relays. In the
second part of this paper, we propose the row-monomial
DOSTBCs-CPI, where the noises at the destination are always
uncorrelated and the relays exploit the CPI of the first hop to
construct the codes. We derive an upper bound of the data-rate of
those codes and find the actual codes achieving this upper bound.
The upper bound of the data-rate of the row-monomial DOSTBC-CPI
suggests that the row-monomial DOSTBCs-CPI have better bandwidth
efficiency than the DOSTBCs and the row-monomial DOSTBCs.
Moreover, the upper bound of the data-rate of the row-monomial
DOSTBC-CPI is independent of the number $K$ of the relays, and
hence, the codes have good bandwidth efficiency even in a
cooperative network with many relays.

Our work can be extended in the following two ways. First, it will
be very interesting to consider a more general case, where the
noises at the destination are possibly correlated and the relays
use the CPI of the first hop to construct the codes. Intuitively,
such codes should have even higher data-rate than the row-monomial
DOSTBCs-CPI. But, we conjecture that the improvement of the
data-rate is just marginal. This is because, by comparing the
DOSTBCs and the row-monomial DOSTBCs, we notice that removing the
row-monomial limitation just slightly improves the data-rate.
Secondly, we can assume that the relays have the full CSI,
including not only the channel phase $\theta_k$ but also the
channel magnitude $|h_k|$, of the first hop and use this
information in the code construction. We notice that the use of
the channel magnitude $|h_k|$ only affects the structure of the
noise covariance matrix $\bmR$; but it can not change the
structure of the code matrix $\bmX_C$. Therefore, we conjecture
that the data-rate can not be improved by assuming the relays have
the full CSI of the first hop.

%%%%%%%%%%%%%%%%%%%%%%%%%%%%%%%%%%%%%%%%%%%%%%%%%%%%%%%%%%%%%%%%%%%%%%%%%%%%%%%

%\clearpage
%\newpage

\appendices
%\vspace{1cm}
\section*{Appendix A}
\renewcommand\theequation{A.\arabic{equation}}
\setcounter{equation}{0}

\begin{center}
Proof of Theorem \ref{thm:necessary}
\end{center}

By following the proof of Lemma \ref{lem:propertyAB}, it can be
easily shown that (\ref{eqn:orthE}) is equivalent with the
conditions (\ref{eqn:Econd1})--(\ref{eqn:Econd5}). On the other
hand, if a
DOSTBC $\bmX_D$ exists, (\ref{eqn:orth}) holds by Definition 1%\ref{defi:DOSTBC}
, and hence, (\ref{eqn:cond1})--(\ref{eqn:cond5}) hold by Lemma
\ref{lem:propertyAB}. Therefore, in order to prove Theorem
\ref{thm:necessary}, we only need to show that, if
(\ref{eqn:cond1})--(\ref{eqn:cond5}) hold,
(\ref{eqn:Econd1})--(\ref{eqn:Econd5}) hold and $E_{n,k}$ is
strictly positive.

We start our proof by evaluating $[\bmR]_{t_1,t_2}$ and
$[\bmR^{-1}]_{t_1,t_2}$. According to (\ref{eqn:R}), when $t_1\neq
t_2$, $[\bmR]_{t_1,t_2}$ can be either null or a sum of several
terms containing $|\rho f_k|^2$; when $t_1 = t_2 = t$,
$[\bmR]_{t,t}$ is a sum of a constant 1, which is from the
identity matrix, and several terms containing $|\rho f_k|^2$.
Therefore, we can rewrite $[\bmR]_{t,t}$ as
$[\bmR]_{t,t}=\bar{R}_{t,t}+1$, where $\bar{R}_{t,t}$ accounts for
all the terms containing $|\rho f_k|^2$. $[\bmR^{-1}]_{t_1,t_2}$
is given by $[\bmR^{-1}]_{t_1,t_2} =
C_{t_2,t_1}/\mathrm{det}(\bmR)$, where $C_{t_2,t_1}$ is the matrix
cofactor of $[\bmR]_{t_2,t_1}$. When $t_1 =t_2=t$, by the
definition of matrix cofactor, $C_{t,t}$ contains a constant 1
generated by the product $\prod_{i=1,i\neq t}^T [\bmR]_{i,i} =
\prod_{i=1,i\neq t}^T (\bar{R}_{i,i}+1)$. Furthermore, it is easy
to see that the constant 1 is the only constant term in $C_{t,t}$.
Thus, $C_{t,t}$ can be rewritten as $C_{t,t} = \bar{C}_{t,t}+1$
and there is no constant term in $\bar{C}_{t,t}$. Consequently,
$[\bmR^{-1}]_{t,t}$ can be rewritten as $[\bmR^{-1}]_{t,t} =
\bar{C}_{t,t}/\mathrm{det}(\bmR) + 1/\mathrm{det}(\bmR)$. When
$t_1 \neq t_2$, $C_{t_2,t_1}$ does not contain any constant term,
and hence, $[\bmR^{-1}]_{t_1,t_2}$ does not contain the term
$1/\mathrm{det}(\bmR)$.\footnote{$C_{t_2,t_1}$ may be zero; but it
does not change the conclusion that $[\bmR^{-1}]_{t_1,t_2}$ does
not contain the term $1/\mathrm{det}(\bmR)$.} Therefore, we can
extract the term $1/\mathrm{det}(\bmR)$ from every main diagonal
entry of $\bmR^{-1}$ and rewrite $\bmR^{-1}$ in the following way
\begin{eqnarray}
\bmR^{-1}
&=&\frac{1}{\mathrm{det}(\bmR)}\bar{\bmC}+\frac{1}{\mathrm{det}(\bmR)}\bmI.
\end{eqnarray}

Then we show that (\ref{eqn:Econd1}) holds if (\ref{eqn:cond1})
holds. If (\ref{eqn:cond1}) holds, we have
\begin{equation}\label{eqn:arazero}
\bmA_{k_1} \bmR^{-1} \bmA_{k_2}^H  =
\frac{1}{\mathrm{det}(\bmR)}\bmA_{k_1} \bar{\bmC} \bmA_{k_2}^H +
\frac{1}{\mathrm{det}(\bmR)} \bmA_{k_1}\bmA_{k_2}^H=\bmzero.
\end{equation}
Note that $\bmR^{-1}$ and $\bar{\bmC}$ are random matrices. In
order to make (\ref{eqn:arazero}) hold for every possible
$\bmR^{-1}$ and $\bar{\bmC}$, both terms in (\ref{eqn:arazero})
must be equal to zero. Therefore, (\ref{eqn:Econd1}) holds.
Similarly, we can show that (\ref{eqn:Econd2})--(\ref{eqn:Econd4})
hold if (\ref{eqn:cond2})--(\ref{eqn:cond4}) hold. Now, we show
that (\ref{eqn:Econd5}) holds if (\ref{eqn:cond5}) holds. If
(\ref{eqn:cond5}) holds, we have
\begin{eqnarray}
\bmA_k \bmR^{-1} \bmA_k^H + \bmB^{*}_k \bmR^{-1} \bmB^T_ k &=&
\frac{1}{\mathrm{det}(\bmR)}\left(\bmA_k \bar{\bmC} \bmA_k^H +
\bmB^{*}_k \bar{\bmC} \bmB^T_ k\right)\nonumber +
\frac{1}{\mathrm{det}(\bmR)}\left(\bmA_k \bmA_k^H +
\bmB^{*}_k\bmB^T_ k
 \right)\\
&=& {\mathrm{diag}}[D_{1,k},\cdots,D_{N,k}].
\end{eqnarray}
For the same reason as in (\ref{eqn:arazero}), the off-diagonal
entries of $\bmA_k \bmA_k^H + \bmB^{*}_k\bmB^T_ k$ must be zero,
and hence, (\ref{eqn:Econd5}) holds.

Lastly, we show that $E_{n,k}$ is strictly positive if
(\ref{eqn:cond5}) holds. From (\ref{eqn:cond5}) and
(\ref{eqn:Econd5}), we have
\begin{eqnarray}
D_{n,k} &=& \sum_{t=1}^T \sum_{i=1}^T [\bmR^{-1}]_{i,t}
([\bmA_k]_{n,i} [\bmA_k]_{n,t}^* + [\bmB_k]_{n,i}^*
[\bmB_k]_{n,t})\\\label{eqn:ARABRAelement2}
 E_{n,k} &=& \sum_{t=1}^T(|[\bmA_k]_{n,t}|^2
+ |[\bmB_k]_{n,t}|^2).\label{eqn:AABBelement2}
\end{eqnarray}
Since $D_{n,k}$ is non-zero, at least one $[\bmA_k]_{n,t}$ or one
$[\bmB_k]_{n,t}$ is non-zero. %Furthermore, the modulus of that
%non-zero entry is 1 by Lemma \ref{thm:propertyAB}.
 Therefore, $E_{n,k}=\sum_{t=1}^T(|[\bmA_k]_{n,t}|^2 + |[\bmB_k]_{n,t}|^2)$ %\geq 1$
  is strictly positive, which completes the proof of Theorem
\ref{thm:necessary}.

\appendices
%\vspace{1cm}
\section*{Appendix B}
\renewcommand\theequation{B.\arabic{equation}}
\setcounter{equation}{0}

\begin{center}
Proof of Theorem \ref{thm:upperDOSTBC}
\end{center}

Let $\underline{\bmA} = [\bmA_1,\cdots,\bmA_K]^T$ and
$\underline{\bmB} = [\bmB_1,\cdots,\bmB_K]^T$; then the dimension
of $\underline{\bmA}$ and $\underline{\bmB}$ is $NK\times T$. From
(\ref{eqn:Econd1}), every row of $\bmA_{k_1}$ is orthogonal with
every row of $\bmA_{k_2}$ when $k_1 \neq k_2$.\footnote{A row
vector $\bmx$ is said to be orthogonal with another row vector
$\bmy$ if $\bmx\bmy^H$ is equal to zero.} Furthermore, because
$\bmA_k$ is column-monomial by Lemma \ref{lem:propertyAB}, every
row of $\bmA_k$ is orthogonal with every other row of $\bmA_k$.
Therefore, any two different rows in $\underline{\bmA}$ are
orthogonal with each other, and hence,
${\mathrm{rank}}(\underline{\bmA}) =
\sum_{k=1}^K{\mathrm{rank}}(\bmA_k)$. Similarly, any two different
rows in $\underline{\bmB}$ are orthogonal with each other, and
hence, ${\mathrm{rank}}(\underline{\bmB}) =\sum_{k=1}^K
{\mathrm{rank}}(\bmB_k)$.

On the other hand, from (\ref{eqn:Econd5}), we have
\begin{equation}
{\mathrm{rank}}(\bmA_k) +{\mathrm{rank}}(\bmB_k) \geq
{\mathrm{rank}}({\mathrm{diag}}[E_{1,k},\cdots,E_{N,k}]) =  N,
\end{equation}
where the inequality is from the rank inequality 3) in
\cite{wang}, and hence,
\begin{eqnarray}
\sum_{k=1}^K {\mathrm{rank}}(\bmA_k) + \sum_{k=1}^K
{\mathrm{rank}}(\bmB_k) \geq NK.
\end{eqnarray}
Because ${\mathrm{rank}}(\underline{\bmA})$ and
${\mathrm{rank}}(\underline{\bmB})$ are integers, we have
\begin{equation}\label{eqn:rankA}
{\mathrm{rank}}(\underline{\bmA}) = \sum_{k=1}^K
{\mathrm{rank}}(\bmA_k) \geq \left\lceil \frac{NK}{2}\right\rceil
\end{equation}
or
\begin{equation}\label{eqn:rankB}
{\mathrm{rank}}(\underline{\bmB}) = \sum_{k=1}^K
{\mathrm{rank}}(\bmB_k) \geq \left\lceil \frac{NK}{2}\right\rceil.
\end{equation}
If (\ref{eqn:rankA}) is true, $T\geq
{\mathrm{rank}}(\underline{\bmA}) \geq \left\lceil
(NK)/2\right\rceil$ and (\ref{eqn:rateDOSTBC}) holds. If
(\ref{eqn:rankB}) is true, the same conclusion can be made.

\appendices
%\vspace{1cm}
\section*{Appendix C}
\renewcommand\theequation{C.\arabic{equation}}
\setcounter{equation}{0}

\begin{center}
Proof of Theorem \ref{thm:partition}
\end{center}

The sufficient part is easy to verify. Thus, we focus on the
necessary part, i.e. if $\bmX_C$ is a row-monomial DOSTBC-CPI, all
the sub-matrices $\tilde{\bmX}_{Cw}$ are also row-monomial
DOSTBCs-CPI. Assume that the dimension of $\bmR_w$ is $T_w\times
T_w$.

Firstly, we show that $\tilde{\bmX}_{Cw}\bmR_w\tilde{\bmX}_{Cw}^H$
is a diagonal matrix. Based on (\ref{eqn:orthcsi}), when $k_1\neq
k_2$, $[\bmX_C\bmR^{-1}\bmX_C^H]_{k_1,k_2}$ is given by
\begin{equation}\label{eqn:xrxk1k1}
[\bmX_C\bmR^{-1}\bmX_C^H]_{k_1,k_2}=\sum_{w=1}^W\sum_{t=1}^{T_w}[\bmX_{Cw}]_{k_1,t}
[\bmX_{Cw}]_{k_2,t}^*R_w = 0.
\end{equation}
If all the terms in this summation are zero, it is trivial to show
that $\sum_{t=1}^{T_w}[\bmX_{Cw}]_{k_1,t} [\bmX_{Cw}]_{k_2,t}^*R_w
= 0$ for $1\leq w\leq W$. Because $\tilde{\bmX}_{Cw}$ contains all
the non-zero entries of $\bmX_{Cw}$, we have
$[\tilde{\bmX}_{Cw}\bmR_w\tilde{\bmX}_{Cw}^H]_{k_1,k_2}
=[\bmX_{Cw}\bmR_w\bmX_{Cw}^H]_{k_1,k_2}=\sum_{t=1}^{T_w}[\bmX_{Cw}]_{k_1,t}
[\bmX_{Cw}]_{k_2,t}^*R_w= 0$, which means
$\tilde{\bmX}_{Cw}\bmR_w\tilde{\bmX}_{Cw}^H$ is a diagonal matrix.

If there is one term $[\bmX_{Cw_1}]_{k_1,t_1}
[\bmX_{Cw_1}]_{k_2,t_1}^*R_{w_1}\neq 0$, some other terms must
cancel this term in order to make (\ref{eqn:xrxk1k1}) hold.
Actually, the non-zero term $[\bmX_{Cw_1}]_{k_1,t_1}
[\bmX_{Cw_1}]_{k_2,t_1}^*R_{w_1}$ must be cancelled by exactly one
other term. This can be shown by contradiction. We assume that
$[\bmX_{Cw_1}]_{k_1,t_1} [\bmX_{Cw_1}]_{k_2,t_1}^*R_{w_1}$ is
cancelled by two other terms together, i.e.
\begin{equation}
[\bmX_{Cw_1}]_{k_1,t_1} [\bmX_{Cw_1}]_{k_2,t_1}^*R_{w_1}
+[\bmX_{Cw_2}]_{k_1,t_2}
[\bmX_{Cw_2}]_{k_2,t_2}^*R_{w_2}+[\bmX_{Cw_3}]_{k_1,t_3}
[\bmX_{Cw_3}]_{k_2,t_3}^*R_{w_3}=0.
\end{equation}
In order to make this equality hold, one of the following three
equalities must hold: 1) $[\bmX_{Cw_2}]_{k_1,t_2}$ $=\pm
[\bmX_{Cw_1}]_{k_1,t_1}$; 2) $[\bmX_{Cw_3}]_{k_1,t_3}=\pm
[\bmX_{Cw_1}]_{k_1,t_1}$; 3)
$\pm[\bmX_{Cw_2}]_{k_1,t_2}=\pm[\bmX_{Cw_3}]_{k_1,t_3}=[\bmX_{Cw_1}]_{k_2,t_1}^*$.
However, those three equalities all contradict with our assumption
that the covariance matrix $\bmR$ is diagonal. For example, we
assume $[\bmX_{Cw_1}]_{k_1,t_1}=s_n^{w_1}$, $1\leq n\leq N_{w_1}$,
and the equality $[\bmX_{Cw_2}]_{k_1,t_2}=\pm
[\bmX_{Cw_1}]_{k_1,t_1}$ holds. Thus, $[\bmX_{Cw_2}]_{k_1,t_2}=\pm
s_n^{w_1}$ and $s_n^{w_1}$ is transmitted in the $k_1$-th row of
$\bmX_C$ for at least twice. This makes the noise covariance
matrix $\bmR$ non-diagonal, which contradicts with our assumption.
If we assume $[\bmX_{Cw_1}]_{k_1,t_1}
[\bmX_{Cw_1}]_{k_2,t_1}^*R_{w_1}$ is cancelled by more than two
other terms, the same contradiction can be seen similarly. Thus,
$[\bmX_{Cw_1}]_{k_1,t_1} [\bmX_{Cw_1}]_{k_2,t_1}^*R_{w_1}$ is
cancelled by exactly one other term in the summation
(\ref{eqn:xrxk1k1}) and we have
\begin{equation}\label{eqn:twoterms}
[\bmX_{Cw_1}]_{k_1,t_1} [\bmX_{Cw_1}]_{k_2,t_1}^*R_{w_1}
+[\bmX_{Cw_2}]_{k_1,t_2} [\bmX_{Cw_2}]_{k_2,t_2}^*R_{w_2}=0.
\end{equation}
Furthermore, because $R_i\neq R_j$ when $i\neq j$,
(\ref{eqn:twoterms}) also implies that $R_{w_1}=R_{w_2}$ and
$w_1=w_2$. This means that, if one term in the summation
(\ref{eqn:xrxk1k1}) is non-zero, it must be cancelled by exactly
one other term, which is from the same sub-matrix $\bmX_{Cw}$.
Therefore, we have $\sum_{t=1}^{T_w}[\bmX_{Cw}]_{k_1,t}
[\bmX_{Cw}]_{k_2,t}^*R_w = 0$ when $k_1\neq k_2$. Because
$\tilde{\bmX}_{Cw}$ contains all the non-zero entries of
$\bmX_{Cw}$, we have
$[\tilde{\bmX}_{Cw}\bmR_w\tilde{\bmX}_{Cw}^H]_{k_1,k_2}$ $
=[\bmX_{Cw}\bmR_w\bmX_{Cw}^H]_{k_1,k_2}=\sum_{t=1}^{T_w}[\bmX_{Cw}]_{k_1,t}
[\bmX_{Cw}]_{k_2,t}^*R_w= 0$, when $k_1\neq k_2$. Therefore,
$\tilde{\bmX}_{Cw}\bmR_w\tilde{\bmX}_{Cw}^H$ is a diagonal matrix.

Secondly, we show that the information-bearing symbols
$s^w_1,\cdots,s^w_{N_w}$ are contained in every row of
$\tilde{\bmX}_{Cw}$. Because every main diagonal entry of
$\bmR_{w}$ is the same, it follows from (\ref{eqn:R}) that every
column in $\bmX_{Cw}$ has non-zero entries at the same rows.
Therefore, the non-zero rows in $\bmX_{Cw}$ does not contain any
zero entries. Since $\tilde{\bmX}_{Cw}$ contains all the non-zero
rows in $\bmX_{Cw}$, every entry in $\tilde{\bmX}_{Cw}$ is
non-zero. Then we assume that
$[\tilde{\bmX}_{Cw}]_{k_1,t_1}=s_n^w$, $1\leq n\leq N_w$. Because
every entry in $\tilde{\bmX}_{Cw}$ is non-zero, we can find
another non-zero entry $[\tilde{\bmX}_{Cw}]_{k_2,t_1}$, $k_1\neq
k_2$, from the $t_1$-th column of $\tilde{\bmX}_{Cw}$. Thus,
$[\tilde{\bmX}_{Cw}\bmR_w\tilde{\bmX}_{Cw}^H]_{k_1,k_2}$ must
contain the term $[\tilde{\bmX}_{Cw}]_{k_1,t_1}
[\tilde{\bmX}_{Cw}]_{k_2,t_1}^*R_{w}$. Because
$[\tilde{\bmX}_{Cw}\bmR_w\tilde{\bmX}_{Cw}^H]_{k_1,k_2}=0$,
$[\tilde{\bmX}_{Cw}]_{k_1,t_1}
[\tilde{\bmX}_{Cw}]_{k_2,t_1}^*R_{w}$ must be cancelled by another
term and we assume it is $ [\tilde{\bmX}_{Cw}]_{k_1,t_2}
[\tilde{\bmX}_{Cw}]_{k_2,t_2}^*R_{w}$, $t_1\neq t_2$. In order to
make $[\tilde{\bmX}_{Cw}]_{k_1,t_1}
[\tilde{\bmX}_{Cw}]_{k_2,t_1}^*R_{w}+[\tilde{\bmX}_{Cw}]_{k_1,t_2}
[\tilde{\bmX}_{Cw}]_{k_2,t_2}^*R_{w}=0$, we must have
$[\tilde{\bmX}_{Cw}]_{k_1,t_2}=\pm [\tilde{\bmX}_{Cw}]_{k_1,t_1}$
or $[\tilde{\bmX}_{Cw}]_{k_2,t_2}^*=\pm
[\tilde{\bmX}_{Cw}]_{k_1,t_1}$. Due to the row-monomial condition,
$[\tilde{\bmX}_{Cw}]_{k_1,t_2}$ can not be $\pm
[\tilde{\bmX}_{Cw}]_{k_1,t_1}$, and hence, we have
$[\tilde{\bmX}_{Cw}]_{k_2,t_2}=\pm
[\tilde{\bmX}_{Cw}]_{k_1,t_1}^*=\pm s_n^{w*}$. This means that the
$k_2$-th row contains the information-bearing symbol $s_n^w$ as
well. Taking a similar approach, we can show that the
information-bearing symbols $s^w_1,\cdots,s^w_{N_w}$ are contained
in every row of $\tilde{\bmX}_{Cw}$.

Because $\tilde{\bmX}_{Cw}\bmR_w\tilde{\bmX}_{Cw}^H$ is a diagonal
matrix and every row of $\tilde{\bmX}_{Cw}$ contains all the
information-bearing symbols $s^w_1,\cdots,s^w_{N_w}$,
$\tilde{\bmX}_{Cw}\bmR_w\tilde{\bmX}_{Cw}^H$ can be written as
\begin{equation}\label{eqn:xrxm}
\tilde{\bmX}_{Cw}\bmR_w\tilde{\bmX}_{Cw}^H =
|s^w_1|^2\bmM_1+\cdots+|s^w_{N_w}|^2\bmM_{N_w},
\end{equation}
where $\bmM_{n}$ are diagonal and all the main diagonal entries
are non-zero. Note that, if the relays only transmit
$\tilde{\bmX}_{Cw}$ to the destination, $\bmR_w$ is actually the
inverse of the noise covariance matrix at the destination. This is
because $\tilde{\bmX}_{Cw}$ and $\bmR_w$ are obtained after the
same column permutations. Therefore, (\ref{eqn:xrxm}) is
equivalent with (\ref{eqn:orthcsi}). Furthermore, since
$\tilde{\bmX}_{Cw}$ is a sub-matrix of $\bmX_C$, it automatically
satisfies D1.1 and the row-monomial condition. Thus, we conclude
that $\tilde{\bmX}_{Cw}$ satisfies Definition $2$ and it is a
row-monomial DOSTBC-CPI.

\appendices
%\vspace{1cm}
\section*{Appendix D}
\renewcommand\theequation{D.\arabic{equation}}
\setcounter{equation}{0}

\begin{center}
Proof of Theorem \ref{thm:row3}
\end{center}

From Theorem \ref{thm:partition}, every sub-matrix
$\tilde{\bmX}_{Cw}$ is a row-monomial DOSTBC-CPI in variables
$s_1^w,\cdots,s_{N_w}^w$. Furthermore, by (\ref{eqn:orthEcsi}),
every sub-matrix $\tilde{\bmX}_{Cw}$ is also a generalized
orthogonal design. For convenience, we refer to any entry
containing $s_{n_w}^w$ as the $s_{n_w}^w$-entry. Similarly, any
entry containing $s_{n_w}^{w*}$ is referred to as the
$s_{n_w}^{w*}$-entry.

By the row-monomial condition, any row in $\tilde{\bmX}_{Cw}$ can
not contain more than one $s_{n_w}^w$-entry or
$s_{n_w}^{w*}$-entry. Therefore, the data-rate of
$\tilde{\bmX}_{Cw}$ is lower-bounded by $1/2$, which is achieved
when every row contains exactly one $s_{n_w}^w$-entry and one
$s_{n_w}^{w*}$-entry for $1\leq n_w\leq N_w$.

Then we show that the data-rate can not be strictly larger than
$1/2$ by contradiction. Without loss of generality, we assume the
first row of $\tilde{\bmX}_{Cw}$ is
$[s_1^w,\cdots,s_{N_w}^w,s_1^{w*},\cdots,s_{N_w^{'}}^{w*}]$, where
$N_w^{'}<N_w$. Hence, the data-rate of $\tilde{\bmX}_{Cw}$ is
$N_w/(N_w+N_w^{'})$ and it is strictly larger than $1/2$.
Furthermore, because every entry in $\tilde{\bmX}_{Cw}$ is
non-zero, this assumption also means that every row in
$\tilde{\bmX}_{Cw}$ contains exactly $N_w+N_w^{'}$ non-zero
entries. Because $s_{N_w^{'}+1}^{w*},\cdots,s_{N_w}^{w*}$ are not
transmitted by the first row, the second row can not have any
$s_{n_w}^{w}$-entries, $N_w^{'}+1\leq n_w\leq N_w$. This can be
shown by contradiction. For example, if the second row has
$s_{N_w^{'}+1}^{w}$ on the first column, the inner product of the
first and second rows must have the term
$s_1^ws_{N_w^{'}+1}^{w*}$. Because $\tilde{\bmX}_{Cw}$ is a
generalized orthogonal design, the inner product of any two rows
must be zero. In order to cancel the term
$s_1^ws_{N_w^{'}+1}^{w*}$, the first row must have an
$s_{N_w^{'}+1}^{w*}$-entry, which contradicts our assumption.
Thus, the second row can not contain any $s_{n_w}^{w}$-entries,
$N_w^{'}+1\leq n_w\leq N_w$. On the other hand, because the second
row must contain exactly $N_w+N_w^{'}$ non-zero entries, it must
have the $s_{n_w}^w$-entries for $1\leq n_w\leq N_w^{'}$ and the
$s_{n_w}^{w*}$-entries for $1\leq n_w\leq N_w$.

Since $K_w>2$, we can do further investigation on the third row of
$\tilde{\bmX}_{Cw}$. The third row is decided by the first and the
second row jointly. Because the first row does not have
$s_{N_w^{'}+1}^{w*},\cdots,s_{N_w}^{w*}$, the third rows can not
have any $s_{n_w}^{w}$-entries, $N_w^{'}+1\leq n_w\leq N_w$.
Furthermore, because the second row does not have any
$s_{n_w}^{w}$-entries, $N_w^{'}+1\leq n_w\leq N_w$, it can be
easily shown that the third row can not have any
$s_{n_w}^{w*}$-entries, $N_w^{'}+1\leq n_w\leq N_w$. Hence, the
third row can only have the $s_{n_w}^{w}$-entries and the
$s_{n_w}^{w*}$-entries for $1\leq n_w\leq N_w^{'}$. There are at
most $2N_w^{'}$ non-zero entries in the third row and it
contradicts with the fact that every row in $\tilde{\bmX}_{Cw}$
contains exactly $N_w+N_w^{'}$ non-zero entries. This means that
the data-rate of $\tilde{\bmX}_{Cw}$ can not be strictly larger
than $1/2$. Because it has been shown that the data-rate of
$\tilde{\bmX}_{Cw}$ is lower-bounded by $1/2$, we conclude that
the data-rate of $\tilde{\bmX}_{Cw}$ is exactly $1/2$ when
$K_w>2$.

\appendices
%\vspace{1cm}
\section*{Appendix E}
\renewcommand\theequation{E.\arabic{equation}}
\setcounter{equation}{0}

\begin{center}
Proof of Theorem \ref{thm:rateDOSTBCscsi}
\end{center}

Like in Theorems \ref{thm:partition} and \ref{thm:row3}, we still
partition $\bmX_C$ into $\bmX_C =[\bmX_{C1},\cdots,\bmX_{CW}]$.
Let $\bmX_C^{k}$ denote the matrix containing all the sub-matrices
$\bmX_{Cw}$ with $k$ non-zero rows, and hence,
$\bmX_C=[\bmX_C^1,\cdots,\bmX_C^K]$. Furthermore, assume the total
number of non-zero entries in $\bmX^{k}_{C}$ is $P_{k}$, and
hence, $\sum_{k=1}^K P_k$ is the total number of non-zero entries
in $\bmX_C$. For convenience, we refer to any entry containing
$s_{n}$ as the $s_{n}$-entry. Similarly, any entry containing
$s_{n}^{*}$ is referred to as the $s_{n}^{*}$-entry.

In order to derive the upper bound of the data-rate, we first
consider the case that $K=3$. For this case, $\bmX_C^{3}$ contains
at most one sub-matrix and we assume $\bmX_C^{3}=\bmX_{C1}$. Thus,
$\bmX_C^{3}$ is a row-monomial DOSTBC-CPI and its data-rate is
exactly $1/2$ by Theorem \ref{thm:row3}. Furthermore, we assume
$\bmX_C^{3}$ is in variables $s_1,\cdots,s_{N_1}$, $1\leq N_1\leq
N$. By the proof of Theorem \ref{thm:row3}, every row of
$\bmX_C^{3}$ contains exactly one $s_n$-entry and one
$s_n^*$-entry, $1\leq n\leq N_1$. Therefore, there is no
$s_n$-entry or $s_n^*$-entry in $\bmX_C^{1}$ and $\bmX_C^{2}$,
$1\leq n\leq N_1$; otherwise, there will be two $s_n$-entries or
two $s_n^*$-entries in a row of $\bmX_C$, which will make the
noise covariance matrix $\bmR$ non-diagonal. Thus, the matrix
$[\bmX_C^1,\bmX_C^2]$ is actually a row-monomial DOSTBC-CPI in
variables $s_{n+1},\cdots,s_N$. Furthermore, because every column
in the matrix $[\bmX_C^1,\bmX_C^2]$ has at most two non-zero
entries, it is easy to show that its data-rate can not be larger
than $1/2$ by following the proof of Theorem $2$ in
\cite{zhihang}. Because the data-rate of $\bmX_C^3$ is exactly
$1/2$ and the data-rate of $[\bmX_C^1,\bmX_C^2]$ is less than
$1/2$, the data-rate of $\bmX_C =[\bmX_{C}^1,\bmX_C^2,\bmX_{C}^3]$
must be upper-bounded by $1/2$ when $K=3$.

Secondly, we consider the case that $K>3$. When $k>2$, the
data-rate of $\bmX_C^{k}$ is exactly $1/2$. This means, if an
information-bearing symbol $s_n$ appears in a row of $\bmX_C^{k}$,
it appears exactly twice. On the other hand, (\ref{eqn:orthEcsi})
implies that every row of $\bmX_C$ must have the
information-bearing symbol $s_n$ for at least once, $1\leq n\leq
N$. Therefore, the following inequality holds
\begin{eqnarray}\label{eqn:NK}
\sum_{k=1}^2 P_{k} + \sum_{k=3}^K \frac{P_{k}}{2} &\geq& NK.
\end{eqnarray}
On the other hand, there are totally $P_{k}/k$ columns in
$\bmX_C^{k}$. Thus, the total number $T$ of columns in
$\bmX_C=[\bmX_C^1,\cdots,\bmX_C^K]$ is given by
\begin{eqnarray}\label{eqn:T}
T &=& \sum_{k=1}^K \frac{P_{k}}{k}.
\end{eqnarray}
By (\ref{eqn:NK}) and (\ref{eqn:T}), it is easy to obtain $2N\leq
T$ under the assumption that $K>3$, and hence, the data-rate of
$\bmX_C$ is upper-bounded by $1/2$ when $K>3$.

%%%%%%%%%%%%%%%%%%%%%%%%%%%%%%%%%%%%%%%%%%%%%%%%%%%%%%%%%%%%%%%%%%%%%%%%

\clearpage
\newpage

%\begin{figure}
%\begin{center}
%\subpostscript{N8K6_error.eps}{0.9\textwidth}
%\end{center}
%\caption{The average Bit Error Rate of the row-monomial
%DOSTBCs-CPI with estimation errors at the relays, $N=8$, $K=6$.}
%\label{fig:n8k6_error}
%\end{figure}


\clearpage
\newpage
\begin{thebibliography}{100}

\bibitem{zhihang} Z. Yi and I.-M. Kim, ``Single-symbol ML
decodable distributed STBCs for cooperative networks,'' {\it IEEE
Trans. Inform. Theory}, vol. 53, pp. 2977--2985, Aug. 2007.


\bibitem{sendonaris1} A. Sendonaris, E. Erkip, and B. Aazhang,
``User cooperation diversity--Part I: System description,''
\emph{IEEE Trans. Commun.}, vol. 51, pp. 1927--1938, Nov. 2003.

\bibitem{sendonaris2} ------, ``User cooperation diversity--Part
II: Implementation aspects and performance analysis,'' \emph{IEEE
Trans. Commun.}, vol. 51, pp. 1939--1948, Nov. 2003.


\bibitem{laneman1} J. N. Laneman, D. N. C. Tse, and G. W. Wornell,
``Cooperative diversity in wireless networks: Efficient protocols
and outage behavior,'' \emph{IEEE Trans. Inform. Theory}, vol. 50,
pp. 3062--3080, Dec. 2004.


\bibitem{laneman2} J. N. Laneman and G. W. Wornell, ``Energy-efficient
antenna sharing and relaying for wireless networks,'' in
\emph{Proc. of WCNC 2000}, vol. 1, Sep. 2000, pp. 7--12.

\bibitem{khan} Md. Z. A. Khan and B. S. Rajan, ``Single-symbol
maximum likelihood decodable linear STBCs,'' \emph{IEEE Trans.
Inform. Theory,} vol. 52, pp. 2062--2091, May 2006.

\bibitem{anghel1} P. A. Anghel and M. Kaveh, ``Exact symbol error
probability of a cooperative network in a rayleigh-fading
environment,'' \emph{IEEE Trans. Wireless Commun.}, vol. 3, pp.
1416--1421, Sep. 2004.

\bibitem{ribeiro} A. Ribeiro, X. Cai, and G. B. Giannakis, ``Symbol error
probabilities for general cooperative links,'' \emph{IEEE Trans.
Wireless Commun.}, vol. 4, pp. 1264--1273, May 2005.

\bibitem{hasna1} M. O. Hasna and M. S. Alouini, ``End-to-end
performance of transmission systems with relays over
rayleigh--fading channels,'' \emph{IEEE Trans. Wireless Commun.},
vol. 2, pp. 1126--1131, Nov. 2003.

\bibitem{hasna2} ------, ``Harmonic mean
and end-to-end performance of
transmission systems with relays,'' \emph{IEEE Trans. Commun.},
vol. 52, pp. 130--135, Jan. 2004.

\bibitem{chen2} D. Chen and J. N. Laneman, ``Modulation and
demodulation for cooperative diversity in wireless systems,''
\emph{IEEE Trans. Wireless Commun.}, vol. 5, pp. 1785--1794, July
2006.

\bibitem{hammerstrom} I. Hammerstr\"{o}m, M. Kuhn and A. Wittneben,
``Impact of relay gain allocation on the performance of
cooperative diversity networks,'' in {\it Proc. IEEE VTC`04}, vol.
3, Sept. 2004, pp. 1815--1819.

\bibitem{zhihang1} Z. Yi and I.-M. Kim, ``Joint optimization of
relay-precoders and decoders with partial channel side information
in cooperative networks,'' {\it IEEE J. Sel. Areas Commun.}, vol.
25, pp. 447--458, Feb. 2007.



\bibitem{zhao1} Y. Zhao, R. Adve, and T. J. Lim, ``Symbol error
rate of selection amplify-and-forward relay systems,'' {\it IEEE
Commun. Lett.}, vol. 10, pp. 757--759, Nov. 2006.

\bibitem{zhao2} ------, ``Improving amplify-and-forward relay
networks: Optimal power allocation versus selection,'' {\it IEEE
Trans. Wireless Commun.}, accepted for publication.

\bibitem{zhihang2} Z. Yi and I.-M. Kim, ``Diversity order analysis of the decode-and-forward
cooperative networks with relay selection,'' {\it IEEE Trans.
Wireless Commun.}, submitted for publication.



\bibitem{laneman3} J. N. Laneman and G. W. Wornell, ``Distributed
space-time-coded protocols for exploiting cooperative diversity in
wireless networks,'' \emph{IEEE Trans. Inform. Theory}, vol. 49,
pp. 2415--2425, Oct. 2003.

\bibitem{nabar} R. U. Nabar, H. B\"{o}lcskei, and F. W.
Kneub\"{u}hler, ``Fading relay channels: Performance limits and
space-time signal designs,'' \emph{IEEE J. Sel. Areas Commun.},
vol. 22, pp. 1099--1109, Aug. 2004.

\bibitem{yang} S. Yang and J.-C. Belfiore, ``Optimal space-time
codes for the MIMO amplify-and-forward cooperative channel,'' {\it
IEEE Trans. Inform. Theory}, vol. 53, pp. 647--663, Feb. 2007.




\bibitem{gamal} H. El Gamal and D. Aktas, ``Distributed space-time filtering for cooperative
wireless networks,'' in {\it Proc. IEEE GLOBECOM'03}, vol. 4, Dec.
2003, pp. 1826--1830.

\bibitem{yiu} S. Yiu, R. Schober, and L. Lampe, ``Distributed space-time block coding,'' {\it IEEE Trans. Commun.},
vol. 54, pp. 1195--1206, July 2006.


\bibitem{murugan} A. Murugan, K. Azarian and H. El Gamal, ``Cooperative lattice coding and
decoding,'' {\it IEEE J. Sel. Areas Commun.}, vol. 25, pp.
268--279, Feb. 2007.

\bibitem{jing1} Y. Jing and B. Hassibi, ``Distributed space-time coding in wireless relay
networks,'' \emph{IEEE Trans. Wireless Commun.}, vol. 5, pp.
3524--3536, Dec. 2006.

\bibitem{li} Y. Li and X.-G. Xia, ``A family of distributed space-time trellis codes
with asynchronous cooperative diversity,'' {\it IEEE Trans.
Commun.}, vol. 55, pp. 790--800, April 2007.


\bibitem{damen} M. O. Damen and A. R. Hammons Jr., ``On distributed space-time
coding,'' in {\it Proc. IEEE WCNC'07}, Mar. 2007, pp. 552--557.

\bibitem{kiran} T. Kiran and B. S. Rajan, ``Partially-coherent distributed
space-time codes with differential encoder and decoder,'' {\it
IEEE J. Sel. Areas Commun.}, vol. 25, pp. 426--433, Feb. 2007.



\bibitem{hua} Y. Hua, Y. Mei, and Y. Chang, ``Wireless antennas-making wireless
communications perform like wireline communications,'' in
\emph{Proc. IEEE AP-S Topical Conference on Wireless Communication
Technology}, Oct. 2003, pp. 47--73.



\bibitem{rajan} G. S. Rajan and B. S. Rajan, ``Distributed space-time codes for
cooperative networks with partial CSI,'' in {\it Proc. IEEE
WCNC'07}, Mar. 2007, pp. 902--906.

\bibitem{jing3} Y. Jing and H. Jafarkhani, ``Using orthogonal and quasi-orthogonal
designs in wireless relay networks,'' {\it IEEE Trans. Inform.
Theory}, accepted for publication, July, 2007.





\bibitem{su3} W. Su and X.-G. Xia, ``On space-time block codes from complex orthogonal
designs,'' {\it Wireless Personal Commun.}, vol. 25, pp. 1--26,
Apr. 2003.


\bibitem{tarokh} V. Tarokh, H. Jafarkhani, and A. R. Calderbank,
``Space-time block codes from orthogonal designs,'' \emph{IEEE
Trans. Inform. Theory,} vol. 45, pp. 1456--1467, July 1999.


\bibitem{wang} H. Wang and X.-G. Xia, ``Upper bounds of rates of complex orthogonal
space-time block codes,'' {\it IEEE Trans. Inform. Theory,} vol.
49. pp. 2788--2796, Oct. 2003.

\bibitem{elia} P. Elia and P. V. Kumar, ``Approximately universal optimality over
several dynamic and non-dynamic cooperative diversity schemes for
wireless networks,'' {\it IEEE Trans. Inform. Theory,} submitted
for publication, Dec. 2005.


\bibitem{alamouti} S. Alamouti, ``A simple transmit diversity technique
for wireless communications,'' \emph{IEEE J. Sel. Areas Commun.},
vol. 16, pp. 1451--1458, Aug. 1998.

%\bibitem{liang2} X.-B. Liang and X.-G. Xia, ``On the nonexistence
%of rate-one generalized complex orthogonal designs,'' {\it IEEE
%Trans. Inform. Theory,} vol. 49, pp. 2984--2989, Nov. 2003.












\end{thebibliography}
\end{document}